# Can Large Language Models Replace Human Coders? Introducing ContentBench

## Michael Haman

Department of Humanities, Faculty of Economics and Management, Czech University of Life Sciences Prague, Czech Republic

Correspondence: haman@pef.czu.cz

**Abstract**: Can low-cost large language models (LLMs) take over the interpretive coding work that still anchors much of empirical content analysis? This paper introduces ContentBench, a public benchmark suite that helps answer this replacement question by tracking how much agreement low-cost LLMs achieve and what they cost on the same interpretive coding tasks. The suite uses versioned tracks that invite researchers to contribute new benchmark datasets. I report results from the first track, ContentBench–ResearchTalk v1.0: 1,000 synthetic, social-media-style posts about academic research labeled into five categories spanning praise, critique, sarcasm, questions, and procedural remarks. Reference labels are assigned only when three state-of-the-art reasoning models (GPT-5, Gemini 2.5 Pro, and Claude Opus 4.1) agree unanimously, and all final labels are checked by the author as a quality-control audit. Among the 59 evaluated models, the best low-cost LLMs reach roughly 97–99% agreement with these jury labels, far above GPT-3.5 Turbo, the model behind early ChatGPT and the initial wave of LLM-based text annotation. Several top models can code 50,000 posts for only a few dollars, pushing large-scale interpretive coding from a labor bottleneck toward questions of validation, reporting, and governance. At the same time, small open-weight models that run locally still struggle on sarcasm-heavy items (for example, Llama 3.2 3B reaches only 4% agreement on hard-sarcasm). ContentBench is released with data, documentation, and an interactive quiz at contentbench.github.io to support comparable evaluations over time and to invite community extensions.

**Keywords**: content analysis; large language models; text annotation; coding; reliability; validity; computational social science

## Introduction

Content analysis is foundational to empirical social science. It is the method through which researchers connect texts to concepts, arguments, and social processes. Traditionally, this work has required teams of trained human coders who read texts, apply interpretive categories, negotiate disagreements, and produce the coded datasets that underpin published findings. This labor is expensive, slow, and difficult to scale, which has long constrained what questions could be asked of large textual datasets.

Large language models (LLMs) raise a fundamental question: can they replace human coders? An LLM that costs fractions of a cent per classification can now do in seconds what a human coder does in minutes. If these models achieve acceptable agreement on interpretive categories, the economic and practical case for human coding weakens considerably.

This paper introduces ContentBench, a benchmark suite designed to help answer the replacement question by providing stable tasks, locked prompts, and transparent evaluation, responding to calls for domain-specific benchmarks aligned with the priorities of social science research (Lin & Zhang, 2025). I report results from the first track, ContentBench–ResearchTalk v1.0, which focuses on coding short posts about academic research into five categories. This is the beginning of a larger project; future tracks will extend to additional domains and languages. My goal is not to definitively resolve whether LLMs should replace human coders, but to provide a public benchmark with initial evidence and invite the research community to test their own coders and models against these results.

The paper addresses three questions. First, what agreement levels can current low-cost models achieve on interpretive classification tasks? Second, how do these results compare to GPT-3.5 Turbo, the model behind early ChatGPT? Third, what does the cost–agreement frontier look like for researchers considering LLM coding at scale?

ContentBench treats "replacement" as a suite-level research question rather than a claim that any single dataset can settle. Each track specifies its own reference-labeling procedure and documents what kind of standard it provides. ContentBench–ResearchTalk v1.0 uses a conservative three-model jury with an author audit to create a stable, high-consensus reference for benchmarking low-cost models; future tracks can and should include traditional human-coded reference labels where appropriate. Crucially, ContentBench does not fix a single notion of "gold labels"; tracks may use expert panels, multi-human coding, crowdsourcing, adjudication, or model-based procedures, as long as the reference protocol is documented.

**Background and Related Work**

The prospect of using large language models (LLMs) as coders in content analysis sits at the intersection of two established problems: how to treat coding as measurement, and how to scale that measurement affordably. Content analysis handbooks set out standards for establishing validity and reliability, including intercoder reliability (Krippendorff, 2019; Neuendorf, 2017), and related work has explored lower-cost annotation infrastructures such as crowdsourcing when expert coding is impractical (Benoit et al., 2016). LLMs intensify these pressures because they assign labels at low marginal cost compared with human teams.

The methodological question is therefore not simply whether LLMs can code, but under which protocols, for which constructs, and how to evaluate them in ways that support valid and reproducible inference. Existing empirical work paints a varied picture: on classification tasks, zero-shot LLMs achieve fair agreement with humans but do not outperform dedicated task-specific models, and reported scores vary widely, including very low results; together with open questions about reproducibility, this evidence suggests caution in treating LLMs as substitutes for human annotation (Ollion et al., 2023; Ziems et al., 2024). This section synthesizes that literature along four lines: the evidence base on LLM coding performance, the measurement foundations that any evaluation presupposes, the hybrid and ensemble designs that address cost and error jointly, and the benchmarking gap that ContentBench is designed to fill.

*Evidence on LLM Coding Performance*

Multiple evaluations report that prompted LLMs can match or exceed crowd and expert labels on structured classification tasks. Comparing performance on four datasets, ChatGPT outperformed crowd workers on stance, relevance, topic, and frame labeling at roughly one-thirtieth of the per-item cost (Gilardi et al., 2023). On political Twitter messages, GPT-4 achieved higher accuracy and reliability than both expert and crowd annotators (Törnberg, 2025), and proved highly accurate when classifying political text across multiple countries and languages, with downstream models trained on its labels performing comparably to those trained on human annotations (Heseltine & Clemm Von Hohenberg, 2024). Beyond political text, few-shot LLMs approach human-coder accuracy on open-ended survey responses across large label spaces (Mellon et al., 2024), and successive GPT versions improve multilingual psychological text classification relative to dictionary-based approaches (Rathje et al., 2024). Earlier work established the cost rationale by showing that GPT-3 pseudo-labels can shrink annotation budgets by 50 to 96 percent without degrading downstream model performance (Wang et al., 2021), and a subsequent multi-task evaluation explored GPT-3's potential as a broadly applicable annotator across NLP tasks (Ding et al., 2022). A parallel stream in qualitative methods uses LLMs less as substitute coders and more as collaborators in thematic exploration, focusing on generating and comparing themes rather than assigning predefined classification labels (Hamilton et al., 2023; Yan et al., 2023).

However, replacement claims prove fragile when extended across tasks, domains, languages, and model generations. Comparative studies that include open-source models and wider task sets find substantial variation, with crowd or expert annotations outperforming LLMs in some settings and newer proprietary models narrowing the gap in others (Kristensen-McLachlan et al., 2025; Stromer-Galley et al., 2025). Off-the-shelf multilingual LLMs perform poorly in low-resource language settings (Bhat & Varma, 2023), specialized legal reasoning resists instruction-only prompting (Thalken et al., 2023), and ethnographic annotation across hundreds of excerpts remains far short of the accuracy needed for dependable automation, even for features on which human coders show strong agreement (Goodall et al., 2026). Sarcasm detection, in particular, yields low reliability for both humans and models, suggesting that certain context-dependent distinctions pose difficulties that are not attributable to model capacity alone (Bojić et al., 2025). Systematic reviews reinforce this picture: performance is mixed, with unresolved questions about reproducibility and English-language primacy (Ollion et al., 2023). Validation studies further demonstrate that common optimization approaches, including prompt refinement, cannot replace evaluation against human-generated labels, while dependence on publicly available benchmark corpora introduces contamination risks that complicate performance claims (Pangakis et al., 2023; Pangakis & Wolken, 2025). Domain-specific assessments reach similar conclusions; in computational frame analysis, for example, generative LLMs fall consistently behind human coders (Kunjar et al., 2025). In social policy debates, GPT-4 Turbo achieves high overall accuracy but precision and recall vary strongly across argument types, leading to a recommendation that human validation remains essential before using automated classifications (Gielens et al., 2025).

A recurring finding across this evidence is that performance depends critically on how the coding protocol is specified. Providing codebook-like definitions, decision rules, and examples improves classification agreement, though results remain sensitive to prompt wording and model selection (Chew et al., 2023; Lupo et al., 2023). Category definitions themselves are a core failure point: underspecified or underrepresented categories can lead to weaker classification performance (Domínguez-Diaz et al., 2025). Recent sociological work extends this insight by proposing that codebooks can be translated into structured prompt architectures, which should be systematically developed and evaluated while attending to systematic misclassification tendencies that risk distorting subsequent analyses (Stuhler et al., 2025).

Stability is a prerequisite for both replicability and substantive interpretation. LLM outputs can vary across repeated runs and minor prompt paraphrases, sometimes falling below reliability thresholds (Reiss, 2023), and formalized prompt stability diagnostics help quantify output variability across semantically similar prompts (Barrie et al., 2024). Implementation details such as batching and item ordering can also affect cheaper models more than stronger ones (Zendel et al., 2024). Fine-tuning can outperform zero-shot prompting in specialized settings (Alizadeh et al., 2025; von der Heyde et al., 2025), but using LLM-generated labels for downstream training can transmit non-random errors and degrade model stability (Lu & Smith, 2025). Because commercial API providers routinely update or deprecate models, results depend on the model version and evaluation date, making protocol documentation essential for any reproducible evaluation.

### *Reliability, Disagreement, and Reference Standards*

Content analysis operationalizes reliability as agreement among independent judges beyond chance. Cohen's kappa provides a chance-corrected coefficient for nominal categories, while Krippendorff's alpha generalizes across coders and measurement levels and remains recommended (Cohen, 1960; Hayes & Krippendorff, 2007; Krippendorff, 2019). In practice, reliability is frequently underreported or assessed with overly permissive metrics (Lombard et al., 2002). For LLM-based coding, a coder is defined by a specific model version, prompt, and sampling configuration; without locking these elements, agreement rates are difficult to interpret and difficult to reproduce.

Yet agreement is also contested as a goal. Qualitative researchers debate intercoder reliability because some question its necessity and appropriateness in qualitative research (O'Connor & Joffe, 2020), and NLP evidence shows that human disagreements about textual inferences persist even with additional context and ratings, motivating evaluation objectives that capture judgment distributions rather than collapse them to single labels (Pavlick & Kwiatkowski, 2019). Even when a human-coded reference standard is used, imperfect validation labels can mislead conclusions about comparative model performance (Song et al., 2020).

These issues are compounded by the social organization of labeling. Annotator demographics and backgrounds influence judgments in systematic ways (Pei & Jurgens, 2023), LLM biases can differ from human biases rather than simply reproducing them (Giorgi et al., 2024), and recent proposals for inclusive annotation advocate deliberately including coders with diverse lived experience to strengthen measurement quality (Kathirgamalingam et al., 2024). For benchmarks that inform replacement decisions, these findings mean that high agreement is necessary but not sufficient: readers also need to know which population or procedure defines the reference standard and whether that standard captures the construct for the intended research setting.

Validity concerns sharpen this point, because LLM coding errors can be systematic and can correlate with covariates of substantive interest. In qualitative interview coding, errors correlate with subject characteristics, producing distorted inferences that carefully produced human annotations are needed to diagnose (Ashwin et al., 2023). In political information settings, audits document only moderate alignment with expert ratings and politically patterned biases that can be amplified by role assignment (Yang & Menczer, 2023). A large-scale audit quantifies this risk by showing that routine configuration decisions such as model selection and prompt wording generate systematic variation that can alter downstream conclusions, enabling both accidental and intentional "LLM hacking" (Baumann et al., 2025). The implication for benchmarking is that evaluation protocols should be locked, documented, and resistant to post-hoc tuning, and that benchmarks should support sensitivity analysis rather than reporting only headline scores.

*Hybrid Designs, Ensembles, and Model-Based Evaluation*

Given the mixed evidence on direct substitution, many contributions focus on hybrid designs that allocate work between humans and models. Uncertainty-guided task allocation distributes labeling effort more efficiently, outperforming random allocation by a substantial margin and framing models as complementary annotators rather than unconditional replacements (Li et al., 2023). Confidence-threshold workflows reduce human effort by escalating only ambiguous cases to human review (Tavakoli & Zamani, 2025). Structured agentic pipelines that decompose coding into prompt generation, classification, and judgment steps, with a human intermediary overseeing the process, can successfully replicate expert codings and extend analysis to multilingual corpora on consumer-grade hardware (Farjam et al., 2025). At the same time, hybrid workflows can alter what counts as a human baseline: experimental evidence shows that providing annotators with LLM-generated suggestions shifts label distributions in subjective tasks and can inflate measured model performance when assisted labels are used for evaluation (Schroeder et al., 2025). Benchmarks should therefore clearly separate reference labels from model outputs.

Beyond workflow design, statistical methods address the problem of imperfect surrogate labels directly. Design-based supervised learning (DSL) combines cheap LLM labels with a smaller set of gold-standard labels to obtain asymptotically unbiased estimates and valid uncertainty quantification (Egami et al., 2023), and cost-aware evaluation derives budget-allocation policies that optimally balance raters of different cost and accuracy (Angelopoulos et al., 2025). Ensemble aggregation offers a complementary route: combining multiple LLMs under principled aggregation criteria can address individual-model weaknesses such as inconsistency and misclassification, strengthening both reliability and classification accuracy (Kamen & Kamen, 2025). Together, these approaches shift the replacement question from which single model should substitute for humans to which combination of models, prompts, repetitions, and human checks yields a defensible measurement instrument at acceptable cost.

The growing use of LLMs as annotation instruments has developed alongside the emergence of LLM-as-judge methods, in which models follow rubrics or instructions to assess outputs at scale (Chiang & Lee, 2023; Li et al., 2024). This is directly relevant when reference labels are produced through model judgment. A central finding of the judge literature is that such reference standards are non-trivial: judges can exhibit bias and internal inconsistency, respond differently to variations in prompt design, and be swayed by superficial artifacts, meaning that minor presentational changes can shift verdicts and rankings (Chen & Goldfarb-Tarrant, 2025; Schroeder & Wood-Doughty, 2024; Wei et al., 2024). Theoretically, scalable evaluation faces hard limits near the frontier: if the judge does not outperform the model being evaluated, no correction procedure can cut ground-truth labeling needs by more than half (Dorner et al., 2024).

Practical responses include multi-model jury aggregation (Chen & Goldfarb-Tarrant, 2025), confidence-informed escalation to stronger models when initial judgments are uncertain, and selective evaluation with formal guarantees on agreement with human annotators (Jung et al., 2024); however, these methods mitigate rather than eliminate the underlying measurement problem, as sensitivity to superficial artifacts remains even under optimized jury configurations

(Chen & Goldfarb-Tarrant, 2025). For benchmarks that use model-based labeling, the implication is that procedures such as juries and human verification should be treated as documented design choices that trade coverage for reliability, so that users can assess what kind of reference the benchmark provides.

*The Benchmarking Gap*

In mainstream AI research, benchmarks have become central infrastructure for tracking model progress and informing research and product decisions, from multitask knowledge benchmarks (Hendrycks et al., 2020) through preference-based evaluation (Chiang et al., 2024) to expert-designed examinations of frontier capabilities (Phan et al., 2025). Social science lacks comparable evaluation tools for the interpretive coding tasks that define much of its empirical workflow.

Prior work establishes that LLM coding can be highly effective, but that performance estimates depend on what is being coded and how the coding instrument is specified. Several questions remain unanswered for replacement decisions in social science practice. Without shared protocols, it is difficult to compare LLM coding results across studies, to disentangle prompt effects from model effects, or to track progress over time, a gap that motivates both best-practice frameworks and multi-model benchmarks (Törnberg, 2024; Ziems et al., 2024). Lin and Zhang (2025) explicitly advocate for developing domain-specific benchmarks that align LLM evaluation with the theoretical and empirical priorities of social science research. Because reference labels vary across projects and disagreement is sometimes substantive rather than noise (O'Connor & Joffe, 2020; Pavlick & Kwiatkowski, 2019), agreement claims are difficult to interpret without a transparent account of how labels were produced (Krippendorff, 2019; Song et al., 2020). Finally, much of the evaluation landscape is not organized around the budget constraints that drive real-world adoption: the relevant question for many researchers is not how the best model performs in absolute terms, but what agreement and error profile can be achieved by low-cost models at scale, and how sensitive that performance is to locked prompts and model versions.

Benchmark projects in the social sciences address parts of this problem by comparing models on shared tasks, including continuous leaderboards across domains and languages (González-Bustamante, 2024b) and comparisons against human-coded gold standards in political content annotation (González-Bustamante, 2024a). ContentBench is designed to complement these efforts by focusing specifically on interpretive classification tasks and by making protocol stability and auditability first-class design goals. It provides versioned tracks with stable splits, a fixed classification prompt, and evaluation that jointly reports agreement and cost for low-cost models. Its labeling procedure uses a documented model jury with subsequent human verification, producing high-consensus reference labels that support reproducible comparison while making explicit that results apply to clearly classifiable cases rather than the full range of borderline interpretations. In this sense, ContentBench aims to turn the heterogeneous LLM-as-coder literature into a comparable measurement infrastructure, allowing researchers to test both human coders and models under the same protocol and to make replacement claims that are commensurable across time and research settings.

**Data and Methods**

ContentBench is designed as a benchmark suite rather than a single dataset, providing a reusable evaluation framework for assessing LLM performance on content analysis tasks across different domains. Each track within the suite represents a domain-specific dataset family, such as ResearchTalk for academic commentary, with planned future tracks covering areas such as political discourse, media coverage, and educational contexts. Each track has versioned releases so that results remain reproducible over time, and within each version there are clearly defined splits for different evaluation purposes.

This architecture serves several goals. It allows the benchmark to evolve as new domains become relevant while maintaining stable identifiers for past evaluations. It separates the question of how well models perform from the question of which specific items were evaluated. And it provides a framework that other researchers can extend with new tracks and splits without disrupting existing comparisons. Current results and documentation are maintained at contentbench.github.io, where leaderboards can be updated as new models become available and prices change.

*Dataset Construction*

ContentBench–ResearchTalk v1.0 consists of synthetic social media posts written in a style resembling online commentary about academic papers. Using real social media data raises significant ethical and legal concerns, including user consent, platform terms of service, GDPR compliance, and potential harm from republishing identifiable content. Synthetic data sidesteps these issues while allowing controlled construction of items across categories. The design prioritizes human legibility and distinctions that matter in applied research.

Posts are generated by two models, GPT-5 and Gemini 2.5 Pro, with an enforced 50/50 split to ensure diversity in generation style. The generation prompt uses an adversarial design: it asks for posts that remain realistic and easy for a human reader to interpret, while being difficult for a standard LLM classifier to label correctly. The generator is instructed to exploit common model weaknesses such as over-reliance on keywords and overly literal interpretation, without making the text unnatural. Each post is constrained to a single salient point, a plausible academic research context, and approximately 80 words with no hashtags, emojis, or special formatting. Sarcasm is operationalized as positive or laudatory language intended as critique, typically by pairing exaggerated praise with weak evidence so that literal wording looks positive while intended meaning is negative. Generated candidates are then filtered through the labeling procedure described below; only posts that pass this filter are retained, ensuring that difficulty comes from interpretive inference rather than label noise.

*Coding Scheme*

The coding scheme has five categories. Sarcastic critique captures comments that nominally praise but contain clear ironic or mocking signals. Genuine critique includes serious, non-sarcastic criticism of methodology, interpretation, or writing. Genuine praise covers sincere positive remarks about a paper's contribution. Neutral query refers to clarifying questions without clear evaluative stance. Procedural statement describes a standard and valid scientific procedure, such as preregistration, power analysis, or covariate control.

Unlike traditional content analysis codebooks, which typically include pages of detailed instructions, examples, and decision rules for human coders, the classification prompt used here is deliberately concise to minimize token usage and maximize cost efficiency at scale. The full prompt is reproduced in Appendix B.

*Reference Labels via Jury Consensus*

Reference labels are defined through a three-model jury of diverse, state-of-the-art reasoning systems: GPT-5, Gemini 2.5 Pro, and Claude Opus 4.1. Each juror independently receives the same classification prompt and returns a single label. A post enters the final dataset only when all three jurors agree unanimously.

This unanimity rule is conservative by design. It discards borderline cases and retains only items where these distinct state-of-the-art models converge. Two of the three jury models also serve as generators in the candidate-production step; because labels are assigned only under three-way unanimity, no single model can determine a label, and any candidate on which the generating model disagrees with the other jurors cannot enter the final dataset. In practice, the jury reached unanimous agreement in 96.5% of evaluated candidates (500 of 518) during construction of the balanced core split, suggesting that the category definitions are sufficiently clear for these models to converge on the same judgment. The resulting labels represent not objective truth but a high-consensus reference that is stable and suitable for evaluating cheaper models on clearly classifiable cases.

After jury labeling, I manually reviewed all 1,000 final item–label pairs as a quality-control step. This author audit checked for obvious mismatches between text and label under the codebook definitions; it is not a substitute for traditional intercoder reliability, but provides a sanity check on the consensus labels while keeping the overall procedure scalable.

### Splits

The track includes two splits. The core split (n=500) contains posts balanced across all five categories (100 per label). For the core split, a generated candidate is accepted only if all three jury models agree and the unanimous jury label matches the generation target. The hard-sarcasm split (n=500) is a challenge subset containing only posts labeled as sarcastic critique, constructed to be difficult for a legacy baseline. For the hard-sarcasm split, I first filter candidate posts to those that GPT-3.5 Turbo misclassifies under the shared classification prompt at generation time, discarding items that GPT-3.5 Turbo labels correctly as sarcastic critique. I then require unanimous jury agreement on the sarcastic critique label. This produces a challenge split concentrated on subtle ironic critique that defeats a legacy baseline in the candidate-generation run while remaining stable under the jury rule. Figure 1 summarizes the dataset construction pipeline for both splits.

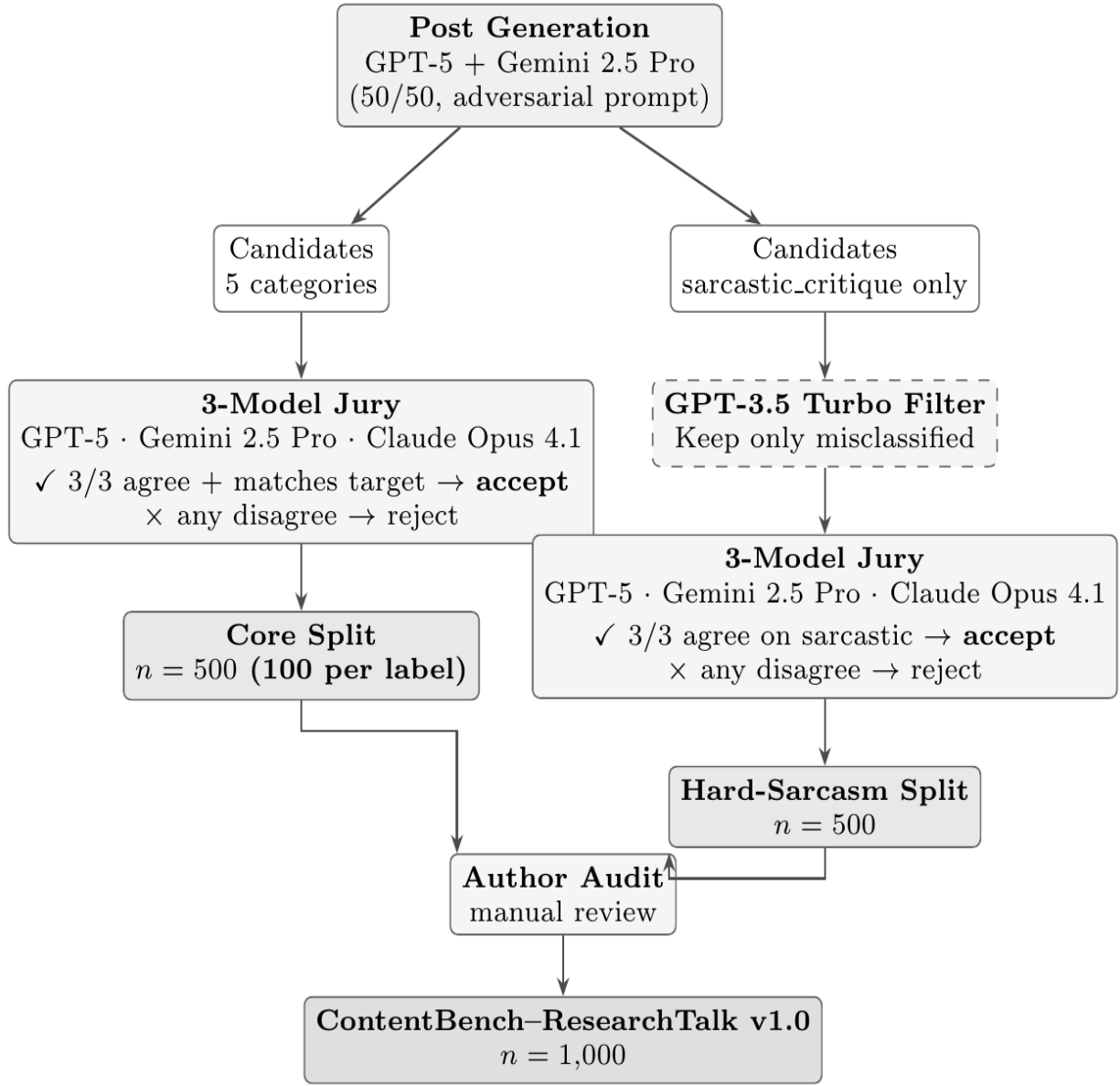

*Figure 1.* Dataset construction pipeline for ContentBench–ResearchTalk v1.0.

*Why Focus on Low-Cost Models?*

A deliberate choice shapes this benchmark: I evaluate low-cost models rather than state-of-the-art reasoning models. State-of-the-art reasoning models certainly perform well, but they are not what most researchers will use at scale.

Content analysis often involves coding thousands or tens of thousands of items. At such volumes, cost matters considerably (Doropoulos et al., 2025). A model that costs substantially more per item may be impractical for a 50,000-post dataset, even if it achieves very high agreement. Early evidence suggested that the trade-off may be favorable: ChatGPT achieved higher accuracy than crowd workers at roughly one-thirtieth of the cost (Gilardi et al., 2023). The cost–performance frontier, not agreement alone, determines what is feasible for applied research (Angelopoulos et al., 2025; Doropoulos et al., 2025).

In practice, many providers offer multiple pricing modes, including on-demand and batch tiers. A model qualifies as "low-cost" if its input price is at or below $0.15 per 1M tokens under a publicly documented pricing tier. In this study, batch tiers were used only where the provider explicitly offers an official batch API with published prices (OpenAI and Google). For models accessed via OpenRouter, I use the token prices listed by OpenRouter at run time.

Eligibility uses the lowest publicly documented tier, while Table 1 reports on-demand prices for comparability; costs use the pricing and billing logs of the service that executed each call. Because model prices change, Table 1 is a fixed snapshot intended for comparability, while the project website provides continuously updated pricing for readers who want to run the benchmark now.

*Model Selection and Evaluation Protocol*

I evaluated 59 models using a shared classification prompt at temperature 0.0 and recorded agreement with the jury labels. Models were accessed via three routes: OpenAI models through the OpenAI API, Google Gemini models through the Gemini API, and a set of open-weight and commercial models (including Llama, Qwen, and others) through OpenRouter. I focused on non-reasoning models because extended thinking token usage makes cost calculation unpredictable, with the exception of OpenAI's GPT-5 series configured for minimal reasoning effort. All API calls reported in this study were made in September 2025.

Because the combined evaluation set is label-imbalanced due to the single-class hard-sarcasm split, I treat overall agreement as a headline metric but not the only diagnostic. On the core split, I report macro-F1 and per-label recall. On the hard-sarcasm split, the minimum informative metric is sarcasm recall, defined as the share of hard-sarcasm items predicted as sarcastic critique.

Costs are calculated from actual billable token usage and pricing snapshots. Unless noted otherwise, cost estimates use on-demand API prices as of September 2025. The classification prompt averages approximately 295 input tokens and 5 output tokens per item, but billable token counts vary across providers due to tokenizers and message-wrapping overhead. I therefore compute costs from provider-reported usage logs rather than relying on a single tokenizer's estimate. For readers who wish to recompute costs under new prices, a useful approximation for 50,000 items is cost ≈ 15 × (input price per million tokens), with output cost typically small in this protocol.

**Results**

Table 1 summarizes results for selected models, combining the core split (n=500) and the hard-sarcasm split (n=500). Agreement indicates match with the jury labels. Pricing in Table 1 is reported as API pricing for comparability; details on the pricing basis are provided in the Methods section.

**Table 1.** Top 10 models on ContentBench–ResearchTalk v1.0.

| Model | Developer | Agreement (%) | Core macro-F1 | Hard-sarcasm recall (%) | $/50k posts |
|---|---|---|---|---|---|
| Gemini 2.5 Flash Preview (09-2025) | Google | 99.8 | 0.996 | 100.0 | 5.10 |
| Gemini 2.5 Flash | Google | 99.6 | 0.992 | 100.0 | 5.10 |
| Gemini 2.5 Flash Lite Preview (09-2025) | Google | 99.4 | 0.992 | 99.6 | 1.59 |
| Gemini 2.0 Flash | Google | 99.2 | 0.984 | 100.0 | 1.57 |
| GPT-5 Mini | OpenAI | 99.0 | 0.990 | 99.0 | 5.03 |
| GLM 4 32B | Zhipu | 98.7 | 0.986 | 98.8 | 1.47 |
| Qwen3 235B A22B Instruct 2507 | Alibaba | 98.4 | 0.970 | 99.8 | 1.51 |
| Llama 4 Maverick | Meta | 98.4 | 0.968 | 100.0 | 2.39 |
| Gemini 2.0 Flash Lite | Google | 98.0 | 0.959 | 100.0 | 1.18 |
| Gemini 1.5 Flash | Google | 97.2 | 0.982 | 96.2 | 1.18 |

*Notes:* Agreement indicates match with jury labels on the combined evaluation set (core + hard-sarcasm). The $/50k column extrapolates per-item cost from the 1,000-item evaluation set to 50,000 posts. Prices are on-demand API rates; details on the pricing basis are provided in the Methods section. Models shown as $0.00 incurred no billable API charges at the time of evaluation (e.g., free-tier access). Full results for all 59 models, split-level metrics, and per-model cost are provided in Appendix A.

Table 2 reports per-label recall on the balanced core split, aggregated across all 59 models in the paper roster. Sarcastic critique is the clear outlier, with mean recall about 0.52 (median about 0.53) versus about 0.93 to 0.96 for the other labels, which motivates the dedicated hard-sarcasm split as a targeted stress test.

**Table 2.** Mean recall by label across 59 models.

| Label | Mean recall | Median recall | IQR (25th-75th) |
|---|---|---|---|
| Genuine critique | 0.956 | 0.980 | 0.950-1.000 |
| Genuine praise | 0.938 | 0.960 | 0.915-0.990 |
| Neutral query | 0.956 | 1.000 | 0.980-1.000 |
| Procedural statement | 0.931 | 0.990 | 0.925-1.000 |
| Sarcastic critique | 0.523 | 0.530 | 0.140-0.925 |

*Notes:* IQR = interquartile range. N = 100 items per label.

To illustrate, two core-split items labeled sarcastic_critique:

"Deeply impressed by Two-Minute Mindfulness Before Exams Improves STEM Performance: A Randomized Classroom Trial. Delivering a statistically significant 0.8-point gain on a 100-point test (p=0.049) with an effect size hovering around d=0.08, and replicating that magnitude across sections, is exactly the kind of robust, scalable impact we need. The authors' discussion rightly highlights the sweeping curricular implications of such a consistently detectable improvement. A model of how marginal gains, carefully celebrated, can meaningfully reshape assessment policy."

"Genuinely thrilled by Smith et al.'s Micro-gestural Synchrony Predicts Team Performance in Hybrid Meetings. The robustness is exceptional: the key association (r=0.19, p=0.048) persists across all specifications, provided participant 12 remains in the sample. The clarity of that boundary condition gives such confidence in generalizability. A masterclass in demonstrating stability by showing the effect disappears only when omitting the single highest-synchrony case. This kind of precision about where your result lives is exactly what the field needs."

Of 59 models, only 11 and 12 respectively classified these items correctly; the rest labeled them genuine_praise.

The top-performing low-cost models achieve 97–99% agreement on the combined evaluation set. GPT-3.5 Turbo achieved 79.6% agreement on the balanced core split under the shared prompt. The hard-sarcasm split is intentionally constructed to isolate subtle ironic critique that defeats GPT-3.5 Turbo in the candidate-generation run, making it a targeted stress test. In contrast, many current low-cost models achieve near-ceiling agreement on these same items, indicating that the baseline for this kind of interpretive stance classification has shifted substantially since the initial wave of LLM-based annotation studies.

These results have implications for interpreting existing research. Studies that concluded LLMs struggle with interpretive coding based on comparisons using GPT-3.5 Turbo or similar systems may have drawn conclusions that no longer hold. At the same time, because ContentBench–ResearchTalk v1.0 is designed to retain only high-consensus cases under a unanimous jury rule, agreement should be interpreted as performance on clearly classifiable items under a locked protocol rather than as a general guarantee about validity across domains, languages, or borderline interpretations.

The cost column illustrates the economics. Under the pricing used here, several top-performing models can code 50,000 posts for on the order of a few dollars. This changes the practical feasibility of large-scale interpretive coding workflows and makes cost–agreement tradeoffs central in methodological decision-making.

**Discussion**

*Implications*

Social scientists increasingly study culture, politics, deviance, discrimination, and institutions through mass digital text. Social media posts, news comments, online reviews, and forum discussions constitute core empirical materials for contemporary social science research. LLM coding makes it feasible to analyze millions of posts at interpretive granularity, applying substantive categories rather than relying solely on word counts or sentiment lexicons. Questions that were previously intractable due to scale become answerable.

Content analysis in the social sciences has developed frameworks for establishing reliability through intercoder agreement, assessing validity through theoretical justification, and ensuring transparency through codebook publication. These frameworks assume human coders. LLM coding does not eliminate the need for such frameworks but demands new equivalents.

Protocol documentation, split-level metrics that expose class-specific failure modes, and explicit cost accounting are practical components of that emerging toolkit, but methodological consensus remains underdeveloped.

*Limitations*

ContentBench–ResearchTalk v1.0 is narrow in scope: short posts, five categories, synthetic data, English only. Performance on other domains, languages, and more complex interpretive tasks may differ substantially. Synthetic data avoids ethical risks of republishing real user content, but it may not capture the full distribution of naturalistic language and context found on real platforms. However, the synthetic design combined with the unanimity requirement offers a complementary measurement advantage. In field evaluation designs, reference labels are typically produced by small coding teams with finite agreement; after reliability is established on a subset of items, remaining items are often coded by a single annotator, and coder rationales are typically unrecorded, making disagreements between model and reference difficult to diagnose (Farjam et al., 2025). Simulation evidence confirms that evaluating automated classifiers against imperfect human annotations carries a substantial risk of incorrect performance conclusions (Song et al., 2020). In this first track, by discarding any item on which three independent jury models do not converge, the benchmark retains only cases where the correct label is unambiguous. Disagreement with these labels is therefore more likely to reflect genuine model failure than reference-standard noise.

Because reference labels in this first track are produced via a model jury, the results should be read as agreement with a documented reference procedure, not as definitive evidence of parity with trained human coding teams; addressing human reference standards is an explicit goal for future tracks in the suite.

A fundamental challenge for reproducibility is that commercial API models are routinely updated, deprecated, or shut down by providers. Any evaluation of API-based models is therefore a time-stamped snapshot of a moving target, which makes versioning and protocol documentation essential for interpretation.

*Open Questions*

The benchmark is motivated by practical methodological uncertainty, not by the claim that one study can settle the replacement debate. I therefore close by posing a set of open questions that the community can use to interpret these results and to guide future benchmark tracks.

Under what conditions, if any, is it acceptable to use LLM coding as the sole basis for content analysis in published research? Are there categories of research questions or coding schemes where LLM-only coding is appropriate, and others where human involvement remains necessary? When reference labels are initially defined by model consensus, even with subsequent human verification, what should validity mean in practice, and what kinds of additional checks are required to justify substantive claims? What reporting standards should methods sections follow when LLM coding is used, given that inconsistent reporting of model versions, prompts, and validation procedures undermines reproducibility? When published findings rest on LLM classifications that later prove systematically biased, how should responsibility and correction be handled, especially when the instrument is a proprietary model that has since changed? Finally, if coding labor shifts from human teams to LLMs, what happens to graduate training in content analysis, and which competencies become essential for credible practice?

**Conclusion**

ContentBench is offered as a benchmark suite designed to inform one practical question: can large language models serve as credible substitutes for human coders in interpretive coding? It evaluates this in content analysis settings by comparing agreement and cost on shared, versioned coding tasks. The first track, ContentBench–ResearchTalk v1.0, is publicly available, and results can be updated over time at contentbench.github.io as new models emerge and prices change.

The evidence reported here is preliminary but notable. Low-cost LLMs now achieve up to 99.8% agreement with high-consensus reference labels on this task, at costs on the order of a few dollars per 50,000 items under the pricing used here. Conclusions drawn from earlier comparisons using systems like GPT-3.5 Turbo may no longer apply to current models.

This is the beginning of a larger project with two main goals. First, future tracks will extend coverage to other domains, languages, and classification tasks. Second, and equally important, the suite will benchmark open-weight models that can run locally on consumer hardware. Given the reproducibility risks of relying on commercial APIs that may be modified or deprecated, researchers need credible local alternatives. Current results show that this goal is not yet achieved for nuanced interpretive categories such as irony detection: Llama 3.2 3B, a model small enough to run on consumer hardware without a dedicated GPU, achieves 76% agreement on the core split but only 4% on hard-sarcasm. Closing this gap is a priority for future work. The concise prompt design used in ContentBench is particularly suited to this goal, since minimal token usage makes local inference more practical on limited hardware.

Researchers who maintain labeled text classification datasets are invited to contribute new tracks to the suite. Each contributed track should include the dataset, classification prompt(s), and documentation of how reference labels were created. The ContentBench dataset and an interactive quiz are publicly available at contentbench.github.io, enabling researchers to test their own human coders and models against the benchmark and to report comparable results. What remains underdeveloped is the methodological and ethical framework for using this capability responsibly. By responding to calls for domain-specific social science benchmarks (Lin & Zhang, 2025) and complementing practical frameworks for integrating LLMs into content analysis workflows (Farjam et al., 2025), I hope this benchmark contributes to that necessary conversation.

**Data Availability**

Project website, benchmark data, code, and full leaderboard: https://contentbench.github.io

# Appendix A: Full Leaderboard

This appendix lists all models in the ResearchTalk v1.0 roster with full coverage on both splits (core and hard-sarcasm) and reports combined agreement, split-level metrics, and cost.

| Rank | Model | Developer | Agreement (%) | Core macro-F1 | Hard-sarcasm recall (%) | $/50k posts |
|---|---|---|---|---|---|---|
| 1 | Gemini 2.5 Flash Preview (09-2025) | Google | 99.8 | 0.996 | 100.0 | 5.10 |
| 2 | Gemini 2.5 Flash | Google | 99.6 | 0.992 | 100.0 | 5.10 |
| 3 | Gemini 2.5 Flash Lite Preview (09-2025) | Google | 99.4 | 0.992 | 99.6 | 1.59 |
| 4 | Gemini 2.0 Flash | Google | 99.2 | 0.984 | 100.0 | 1.57 |
| 5 | GPT-5 Mini | OpenAI | 99.0 | 0.990 | 99.0 | 5.03 |
| 6 | GLM 4 32B | Zhipu | 98.7 | 0.986 | 98.8 | 1.47 |
| 7 | Llama 4 Maverick | Meta | 98.4 | 0.968 | 100.0 | 2.39 |
| 8 | Qwen3 235B A22B Instruct 2507 | Alibaba | 98.4 | 0.970 | 99.8 | 1.51 |
| 9 | Gemini 2.0 Flash Lite | Google | 98.0 | 0.959 | 100.0 | 1.18 |
| 10 | Gemini 1.5 Flash | Google | 97.2 | 0.982 | 96.2 | 1.18 |
| 11 | Gemma 3 27B | Google | 97.1 | 0.976 | 96.6 | 0.00 |
| 12 | Llama 4 Scout | Meta | 96.9 | 0.966 | 97.8 | 1.27 |
| 13 | LongCat Flash Chat | Meituan | 96.6 | 0.990 | 94.2 | 2.51 |
| 14 | Qwen3 Next 80B A3B Instruct | Alibaba | 96.6 | 0.977 | 95.6 | 2.37 |
| 15 | Gemini 2.5 Flash Lite | Google | 96.1 | 0.972 | 95.0 | 1.58 |
| 16 | Qwen2.5 72B Instruct | Alibaba | 96.1 | 0.972 | 95.0 | 1.88 |
| 17 | Llama 3.1 70B Instruct | Meta | 95.2 | 0.976 | 92.8 | 1.54 |
| 18 | Llama 3.3 70B Instruct | Meta | 94.7 | 0.978 | 91.6 | 0.19 |
| 19 | Hermes 3 70B Instruct | Nous Research | 93.8 | 0.976 | 90.0 | 1.85 |
| 20 | Qwen3 30B A3B Instruct 2507 | Alibaba | 91.0 | 0.956 | 86.4 | 1.41 |
| 21 | Gemini 1.5 Flash 8B | Google | 89.0 | 0.918 | 86.2 | 0.59 |
| 22 | Mistral Small 3.2 24B | Mistral | 84.1 | 0.935 | 74.6 | 1.56 |
| 23 | Mistral Small 3.1 24B | Mistral | 82.7 | 0.929 | 72.4 | 1.56 |
| 24 | Mistral 7B Instruct | Mistral | 80.5 | 0.914 | 70.0 | 2.01 |
| 25 | Devstral Small 1.1 | Mistral | 80.3 | 0.922 | 68.2 | 1.11 |
| 26 | Nova Micro 1.0 | Amazon | 76.6 | 0.881 | 64.8 | 0.53 |
| 27 | Devstral Small 2505 | Mistral | 72.0 | 0.904 | 53.2 | 0.91 |
| 28 | Gemma 3 12B | Google | 70.6 | 0.874 | 53.4 | 0.00 |
| 29 | Llama 3.1 8B Instruct | Meta | 69.4 | 0.879 | 53.4 | 0.31 |
| 30 | Mistral 7B Instruct v0.3 | Mistral | 66.5 | 0.867 | 46.4 | 0.46 |

| Rank | Model | Developer | Agreement (%) | Core macro-F1 | Hard-sarcasm recall (%) | $/50k posts |
|---|---|---|---|---|---|---|
| 31 | Mistral Small 3 | Mistral | 65.0 | 0.853 | 43.6 | 0.75 |
| 32 | Nova Lite 1.0 | Amazon | 64.7 | 0.860 | 42.4 | 0.91 |
| 33 | Command R (08-2024) | Cohere | 64.2 | 0.821 | 46.0 | 2.41 |
| 34 | Mistral Nemo | Mistral | 61.2 | 0.844 | 36.8 | 0.30 |
| 35 | Hermes 2 Pro - Llama-3 8B | Nous Research | 59.1 | 0.839 | 33.6 | 0.38 |
| 36 | Gemma 3n 4B | Google | 57.1 | 0.801 | 32.8 | 0.00 |
| 37 | GPT-4o-mini (2024-07-18) | OpenAI | 55.0 | 0.824 | 25.6 | 2.31 |
| 38 | GPT-4o-mini | OpenAI | 54.5 | 0.808 | 25.8 | 2.31 |
| 39 | Phi 4 | Microsoft | 52.9 | 0.737 | 29.2 | 0.91 |
| 40 | Llama 3 8B Instruct | Meta | 52.8 | 0.809 | 23.0 | 0.46 |
| 41 | Qwen-Turbo | Alibaba | 50.8 | 0.800 | 19.0 | 0.79 |
| 42 | Lumimaid v0.2 8B | NeverSleep | 45.0 | 0.686 | 19.6 | 1.49 |
| 43 | ERNIE 4.5 21B A3B | Baidu | 44.2 | 0.717 | 22.4 | 1.23 |
| 44 | Llama 3 8B Lunaris | Sao10K | 42.7 | 0.722 | 10.0 | 0.60 |
| 45 | GPT-5 Nano | OpenAI | 40.5 | 0.717 | 4.6 | 0.98 |
| 46 | Llama 3.2 3B Instruct | Meta | 40.3 | 0.742 | 4.2 | 0.30 |
| 47 | Ministral 8B | Mistral | 39.5 | 0.719 | 2.2 | 1.48 |
| 48 | Qwen2.5 7B Instruct | Alibaba | 39.4 | 0.723 | 0.0 | 0.62 |
| 49 | Phi-3 Mini 128K Instruct | Microsoft | 39.2 | 0.694 | 4.4 | 1.72 |
| 50 | Ministral 3B | Mistral | 39.1 | 0.716 | 0.6 | 0.59 |
| 51 | Llama 3.2 1B Instruct | Meta | 38.2 | 0.673 | 6.0 | 0.08 |
| 52 | LFM 7B | Liquid | 38.1 | 0.692 | 2.6 | 0.16 |
| 53 | Phi-3.5 Mini 128K Instruct | Microsoft | 37.9 | 0.686 | 0.8 | 1.72 |
| 54 | Mistral 7B Instruct v0.1 | Mistral | 37.3 | 0.713 | 12.0 | 1.92 |
| 55 | Gemma 3 4B | Google | 36.3 | 0.667 | 1.0 | 0.00 |
| 56 | AFM 4.5B | Arcee AI | 34.6 | 0.613 | 2.4 | 1.57 |
| 57 | Command R7B (12-2024) | Cohere | 34.5 | 0.628 | 0.0 | 0.60 |
| 58 | UI-TARS 7B | ByteDance | 33.3 | 0.628 | 0.0 | 1.53 |
| 59 | LFM 3B | Liquid | 32.3 | 0.553 | 2.8 | 0.33 |

## Appendix B: Classification Prompt

The {text} placeholder is replaced with each item's post. The role instruction is also passed as the system prompt.

Role:
You are an expert research assistant trained to analyze comments about academic papers. Classify the following post into one of the five categories from the codebook. Provide only the single correct code label and nothing else.

Codebook:
sarcastic_critique: The post uses positive language with an ironic tone to express criticism.
procedural_statement: The post accurately describes a standard and valid scientific procedure.
genuine_praise: The post expresses straightforward and sincere praise for the paper.
genuine_critique: The post presents a direct and non-ironic criticism of the paper.
neutral_query: The post asks a straightforward question for the purpose of clarification.

Post to Classify:
{text}

Your Classification:

**Appendix C: Generation Prompt**

For each generation call, the target category label is appended after "Target Category:" at the end of this prompt.

Role:
You are an advanced creative AI assistant specializing in crafting adversarial examples. Your task is to generate a short, realistic social media post or online comment about a plausibly titled fictional academic paper from any field (quantitative, qualitative, or theoretical).

Primary Objective:
The post you generate MUST be very difficult for a standard AI Language Model to classify correctly, but it should be easy and intuitive for a human expert to understand. The difficulty for the AI should come from its known weaknesses: reliance on keywords, literal interpretation, and lack of contextual or real-world common sense.

Crucial Constraints for Realism:
1. Focus on a SINGLE, Salient Point: Whether it's a critique or praise, the post should revolve around one core idea. Avoid a scattered "laundry list" of issues. A real comment is typically focused.
2. Maintain Plausible Context: The scenario described must be believable within the norms of academic research and discourse. The critique or praise should be for something that *could* realistically appear in a paper, not a parody.
   -For quantitative work: This means using subtly weak numbers (e.g., a low correlation, a borderline p-value, a slightly underpowered sample).
   -For qualitative work: This could mean criticizing a slightly narrow participant pool or praising a particularly insightful piece of analysis.
   -For theoretical work: This might involve pointing out a subtle logical inconsistency or praising a novel conceptual link.
3. Use Authentic Academic Phrasing: The fictional paper's title and the language used in the post should sound like they were written by a real academic. Use standard phrasing and avoid overly simplistic or exaggerated language.
4. Format and Length: Target about 80 words. No hashtags, emojis, or quotation marks. No line breaks.

Your Task:
You will be given a "Target Category" from the codebook below. You must generate a post that is a perfect example of that category, embodying the adversarial principles described above.

Here is the full codebook for your reference:

sarcastic_critique: The post uses positive or laudatory language with clear sarcastic intent to criticize the paper. The key is a disconnect between weak data described and exaggerated praise.
procedural_statement: The post must describe a completely standard and valid scientific practice. The tone must be neutral.

genuine_praise: The post expresses straightforward, non-sarcastic, and contextually plausible praise.
genuine_critique: The post presents a direct, non-sarcastic, and non-adversarial criticism.
neutral_query: The post asks a straightforward, non-rhetorical question for clarification.

Instructions:

Now, generate one social media post that is a perfect example of the following Target Category. Remember to prioritize making it very difficult for another LLM to classify correctly. Your output must ONLY be the content of the post itself, with no extra commentary, titles, or quotation marks.

Target Category: